\documentclass[aps,pra,twocolumn,a4paper,reprint,noeprint]{revtex4-1}
\usepackage[utf8]{inputenc}
\usepackage{graphicx}
\usepackage{amsmath}
\usepackage{amsfonts}
\usepackage{amssymb}
\usepackage{bbm}
\usepackage[usenames,dvipsnames,svgnames,table]{xcolor}
\usepackage[USenglish]{babel}
\usepackage{hyperref}
\usepackage{grffile}
\usepackage{xcolor}
\usepackage{bm}

\newcommand{\ignore}[1]{}

\allowdisplaybreaks

\newcommand{\ket}[1]{\left| #1 \right\rangle}
\newcommand{\bra}[1]{\left\langle #1 \right|}

\begin{document}

\title{Negative longitudinal magnetoconductance at weak fields in Weyl semimetals}

\author{Andy Knoll}
\email{andy.knoll@tu-dresden.de}
\affiliation{Institute of Theoretical Physics, Technische Universit\"at Dresden, 01062 Dresden, Germany}
\author{Carsten Timm}
\email{carsten.timm@tu-dresden.de}
\affiliation{Institute of Theoretical Physics, Technische Universit\"at Dresden, 01062 Dresden, Germany}
\author{Tobias Meng}
\email{tobias.meng@tu-dresden.de}
\affiliation{Institute of Theoretical Physics, Technische Universit\"at Dresden, 01062 Dresden, Germany}

\date{December 20, 2019}

\begin{abstract}
Weyl semimetals are topological materials that provide a condensed-matter realization of the chiral anomaly. A positive longitudinal magnetoconductance quadratic in magnetic field has been promoted as a diagnostic for this anomaly. By solving the Boltzmann equation analytically, we show that the magnetoconductance can become negative in the experimentally relevant semiclassical regime of weak magnetic fields. This effect is due to the simultaneous presence of the Berry phase and the orbital magnetic moment of carriers and occurs for sufficiently strong intervalley scattering. 
\end{abstract}

\maketitle

\textit{Introduction}. Topological states of matter are characterized by topological invariants that can only be detected via non-local probes. The Hall conductance is a prototypical example: It nonlocally relates an electric field to a perpendicular electric current and thus allows one to measure the Chern number of an insulator \cite{Laughlin1981,tknn1982,Kohmoto1985}. While transport reliably detects Chern numbers in gapped systems, the situation is less clear in gapless topological states of matter, such as Weyl semimetals. These are three-dimensional materials in which pairs of otherwise nondegenerate bands touch at isolated points in momentum space \cite{Herring1937,Wan2011,Armitage2018}. These band touching points, also referred to as Weyl nodes, are topological objects. They are protected by a Chern number $\pm 1$ defined on a sphere in momentum space enclosing a given Weyl node. Physically, this Chern number is associated with an anomalous Hall effect. Since the magnitude of this effect depends on the separation between Weyl nodes in momentum space, however, the anomalous Hall effect is not sufficient to distinguish Weyl semimetals from other materials exhibiting anomalous Hall effects. 

The topology of Weyl nodes is also reflected by the so-called chiral anomaly, a characteristic nonconservation of the number of electrons close to a given node under applied electromagnetic fields \cite{Adler1969_a,Adler1969_b,Nielsen1983}. The chiral anomaly has been proposed to lead to a putative smoking-gun signature of Weyl physics in transport, namely, a positive magnetoconductance proportional to the square of the applied magnetic field \cite{SonSpivak2013}. Such a behavior was indeed observed in a number of Weyl semimetals \cite{Huang2015_b,Li2016,Hirschberger2016,Zhang2016_b}.

Later studies, however, have revealed that magnetoconductance measurements can lead to \emph{false positive} results: Materials without Weyl nodes may still exhibit a positive magnetoconductance. First, a positive magnetoconductance was observed for ultraclean PdCoO$_2$ \cite{Kikugawa2016}, which does not host Weyl nodes \cite{KimChoiMin2009,Noh2009}. Second, current jetting effects in topologically trivial systems with nonuniform current distribution can also cause a positive magnetoconductance \cite{Reis2016}. Third, generic metals in the quantum limit of very strong magnetic fields may also exhibit a positive magnetoconductance~\cite{Goswami2015}.

It is thus an important question under which conditions transport is indicative of Weyl physics. At very strong magnetic fields, the answer depends on the dominating scattering mechanism of the system. For short-range point impurities, the longitudinal magnetoconductance can be either negative if the dependence of the Fermi velocity on the magnetic field is included \cite{Goswami2015,Lu2015,Chen2016}, or constant if the Fermi velocity does not depend on the magnetic field \cite{Zhang2016,Shao2019}. Gaussian impurities usually cause a positive linear longitudinal magnetoconductance \cite{LiRoyDasSarma2016,Goswami2015,Shao2019,Zhang2016} and in the presence of screened but not pointlike Coulomb scattering, the longitudinal magnetoconductance is expected to be positive and quadratic~\cite{LiRoyDasSarma2016,Goswami2015,Shao2019,JiLu2018}.  

For the experimentally most relevant weak-field limit, most authors agree on a positive quadratic longitudinal magnetoconductance for isotropic Weyl nodes \cite{SpivakAndreev2016,SonSpivak2013,DasAgarwal2019,Imran2018,Kim2014,Dantas2018,Johansson2019,Burkov2014,PesinMishchenko2015}, even when an orbital magnetic moment (OMM) is included \cite{DasAgarwal2019OMM,GrushinVenderbos2016,Cortijo2016}. In contrast, Zyuzin \cite{Zyuzin2017} finds a longitudinal magnetoconductance proportional to the magnetic field to the power $3/2$ at high temperatures but agrees on the positive sign. On the other hand, Johansson \textit{et al.}\ \cite{Johansson2019} show that anisotropic Weyl nodes or misalignment between the electric and magnetic field can, in principle, cause a negative longitudinal magnetoconductance. 

In this Rapid Communication, we show that the magnetoconductance of Weyl semimetals is not necessarily positive even in the case of isotropic Weyl nodes subject to weak magnetic fields. This conclusion derives from an analytical solution of the Boltzmann equation for nonmagnetic point scatterers. We find that the sign of the magnetoconductance depends on the relative strength of internode and intranode scattering when the OMM is taken into account. 

\begin{figure}
	\centering
	\includegraphics{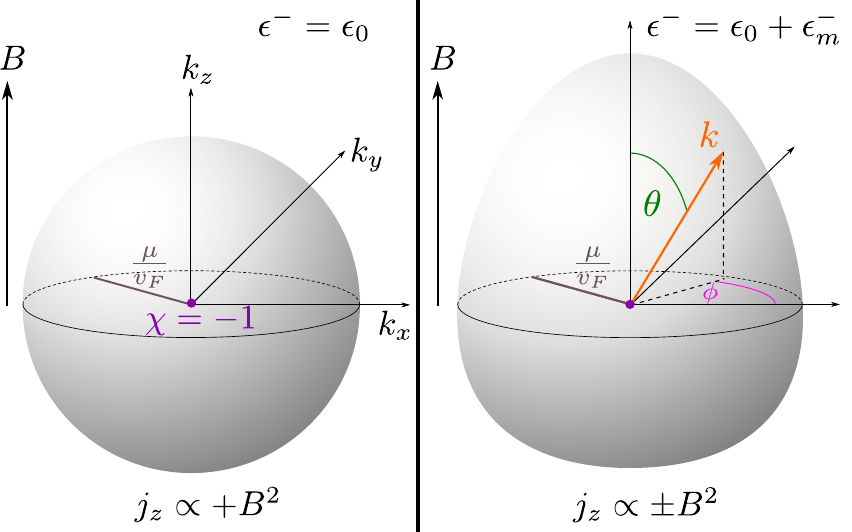}
	\caption{Sketch of the Fermi surface surrounding a Weyl node if the OMM is not (left) or is (right) taken into account. The inclusion of the OMM leads to an egg-shaped Fermi surface, and the magnetoconductance can be either positive or negative, depending on the scattering amplitudes.}
	\label{fig_1}
\end{figure}

\textit{Model}. In order to focus on generic low-energy physics, we study a minimal model of two isotropic Weyl nodes of opposite chirality. We take into account nonmagnetic point scatterers, but strip the model of any nonuniversal complications. We assume the Weyl nodes to reside at the same energy, which we take to be zero. For magnetic field $\bm{B}\rightarrow \bm{0}$, the Hamiltonian in the vicinity of the Weyl node of chirality $\chi=\pm$ takes the form 
\begin{equation}
\label{def_WeylHamiltonian}
H^\chi(\bm{k}) = \chi\, v_F\,\bm{k}\cdot\bm{\sigma},
\end{equation}
where $v_F$ is the Fermi velocity, $\bm{k}$ denotes the momentum measured relative to the node, and $\bm{\sigma}$ is the vector of Pauli matrices. We set $\hbar=c=1$ throughout this Rapid Communication. Since we consider point scattering, the precise positions of the two Weyl nodes in momentum space are irrelevant for our results.

We use Boltzmann theory to study transport in experimentally relevant weak electric and magnetic fields. This semiclassical limit requires the Landau level splitting to be negligible and thus that $B \ll 1/e\: (\mu/v_F)^2$ holds, where $\mu$ is the chemical potential \footnote{See Supplemental Material at \href{http://link.aps.org/supplemental/10.1103/PhysRevB.101.201402}{http://link.aps.org/supp-lemental/10.1103/PhysRevB.101.201402} for details on the semiclassical limit and the vector mean free path.}. Without loss of generality, we consider positive chemical potentials $\mu>0$, in which case only the upper band with the dispersion $\epsilon_0(\bm{k})=+v_F\,k$ is relevant at low temperatures. For this band, the Bloch vectors $\ket{u^\chi (\bm{k})}$ in the vicinity of the Weyl node with chirality $\chi$ are
\begin{align}
	\ket{u^+(\bm{k})}&=\cos \frac{\theta}{2}\: e^{-i\phi}\, \ket{\uparrow}
	  + \sin\frac{\theta}{2}\, \ket{\downarrow}, \\
	\ket{u^-(\bm{k})}&=-\sin\frac{\theta}{2}\: e^{-i\phi}\, \ket{\uparrow}
	  + \cos\frac{\theta}{2}\, \ket{\downarrow}, 
\end{align}
where $\{\ket{\uparrow},\ket{\downarrow}\}$ is the $\sigma_z$ eigenbasis. The polar and azimuthal angles of $\bm{k}$, $\theta$ and $\phi$, respectively, are indicated in Fig.\ \ref{fig_1}. These Bloch vectors relate to two key ingredients in our semiclassical transport theory: The effect of the chiral anomaly is encoded in the Berry curvature,
\begin{equation}
\label{def_Berrycurve}
	\bm{\Omega}^\chi_{\bm{k}}
	= i\bm{\nabla}_{\bm{k}}\times\bra{u^\chi (\bm{k})}\bm{\nabla}_{\bm{k}}\ket{u^\chi (\bm{k})}
	= -\frac{\chi}{2}\, \frac{\bm{k}}{k^3}.
\end{equation} 
In addition, the Bloch vectors also define an OMM. The OMM corresponds to a self-rotation of the wave packets that form the basis of semiclassical transport theory \cite{Xiao2010,SundamNiu1999} and reads
\begin{align}
\label{def_OMM}
	\bm{m}^\chi_{\bm{k}} &= e\: \mathrm{Im} \bra{\frac{\partial u^\chi}{\partial \bm{k}}} 
	\times \left[\epsilon_0(\bm{k})-\hat{H}^\chi(\bm{k})\right]\ket{\frac{\partial u^\chi}{\partial \bm{k}}}
	\nonumber\\ 
	&= -\frac{\chi e v_F}{2}\, \frac{\bm{k}}{k^2},
\end{align}
where $e > 0$ is the elementary charge. The coupling between OMM and magnetic field gives rise to an energy $\epsilon^\chi_m(\bm{k})=-\bm{m}^\chi_{\bm{k}}\cdot \bm{B}$ so that the electron dispersion becomes $\epsilon^\chi(\bm{k})=\epsilon_0(\bm{k})+\epsilon^\chi_m(\bm{k})$. For a magnetic field applied along the $z$ direction, $\bm{B}=B\bm{e}_z$, the OMM energy leads to egg-shaped Fermi surfaces, as illustrated in Fig.\ \ref{fig_1}: The Fermi momentum depends on the angle $\theta$ and the chirality of the Weyl node as $k^\chi(\theta) = (\mu/2v_F)\,(1+\sqrt{1-\chi 4\alpha\cos\theta})$, where $\alpha=eBv_F^2/2\mu^2$.

\textit{Boltzmann formalism}. It is known that the inclusion of the OMM can, in principle, affect transport and other properties in Weyl semimetals \cite{Zyuzin2018,ZhouChang2018,Morimoto2016,McCormick2017,ZyuzinZyuzin2017,DasAgarwal2019OMM,GrushinVenderbos2016,Cortijo2016,Zyuzin2017,ChangYang2015,Pellegrino2015}. Here, we study its impact on magnetotransport via an exact solution of the Boltzmann equation, 
\begin{equation}
\label{def_Boltzmanneq}
	\frac{\partial f^\chi}{\partial t} + \dot{\bm{r}}^\chi\cdot\bm{\nabla}_{\bm{r}} f^\chi
	+ \dot{\bm{k}}^\chi\cdot\bm{\nabla}_{\bm{k}}f^\chi = I_\mathrm{col}\left[f^\chi\right],
\end{equation}
which determines the non-equilibrium distribution function $f^\chi=f^\chi_{\bm{k}}(\bm{r},t)$ of electrons close to the Weyl node of chirality $\chi$ through the equations of motion~\cite{Xiao2010,SundamNiu1999},
\begin{align}
	\dot{\bm{r}}^\chi &= \bm{v}^\chi_{\bm{k}} -\dot{\bm{k}}^\chi\times\bm{\Omega}^\chi_{\bm{k}},
	\label{def_EOM_r}\\
	\dot{\bm{k}}^\chi &= -e\left(\bm{E}+\dot{\bm{r}}^\chi\times\bm{B}\right),
	\label{def_EOM_k}
\end{align}
where $v_{\bm{k}}^\chi=\bm{\nabla}_{\bm{k}}\epsilon^\chi$ is the group velocity. The last term in Eq.\ \eqref{def_EOM_r} is an anomalous velocity, in which the Berry curvature acts as a magnetic field in momentum space.

The collision integral $I_\mathrm{col}[f^\chi]$ describes the scattering processes relaxing the nonequilibrium distribution towards the Fermi function. For elastic, nonmagnetic point scattering, the collision integral reads as
\begin{equation}
\label{def_collisionintegral}
	I_\mathrm{col}\left[f^\chi\right] = \sum_{\chi^\prime,\bm{k}^\prime}
	W^{\chi \chi^\prime}_{\bm{k}\bm{k}^\prime}\left(f^{\chi^\prime}_{\bm{k}^\prime}-f^{\chi}_{\bm{k}}\right).
\end{equation}  
The scattering rate $W^{\chi\chi^\prime}_{\bm{k}\bm{k}^\prime}$ is calculated using Fermi's golden rule,
\begin{align}
\label{def_scatteringrate}
	W^{\chi \chi^\prime}_{\bm{k}\bm{k}^\prime}
	= &2\pi\, \frac{n}{\cal{V}} \left| \big< u^{\chi^\prime}(\bm{k}^\prime)\big|
	\hat{V}^{\chi\chi^\prime}_{\bm{k}\bm{k}^\prime}
	\big|u^{\chi}(\bm{k})\big> \right|^2\nonumber\\
&\times\delta\boldsymbol{\left(\right.}\!\!\epsilon^{\chi^\prime}(\bm{k}^\prime)-\epsilon^{\chi}(\bm{k})\!\!\boldsymbol{\left.\right)},
\end{align}
where the impurity concentration is $n$, $\cal{V}$ denotes the system volume, and the scattering matrix elements are $\hat{V}^{\chi\chi^\prime}_{\bm{k}\bm{k}^\prime} = \mathbbm{1}\, V^{\chi\chi^\prime}$. In some previous works \cite{SonSpivak2013,Kim2014,Lundgren2014}, the collision integral has been described using the relaxation-time approximation. Moreover, the intervalley scattering amplitude $V_\mathrm{inter}$ has been assumed to be much smaller than the intravalley scattering amplitude $V_\mathrm{intra}$, which is appropriate if long-range scattering dominates. Johansson \textit{et al.}\ \cite{Johansson2019} consider uncorrelated point scatterers, which corresponds to $V_\mathrm{inter} = V_\mathrm{intra}$. (They also go beyond the relaxation-time approximation by including in-scattering terms but the scattering integral reduces to the relaxation-time form for their case of point scattering.) Here, we not only include in-scattering terms but also allow for arbitrary intra- and intervalley scattering amplitudes and, most importantly, include the OMM neglected by Johansson \textit{et al}.~\cite{Johansson2019}.

To solve the Boltzmann equation, we focus on the stationary state, a uniform system, and the zero-tem\-pe\-ra\-ture limit. The distribution function then does not depend on time and position. We furthermore write the nonequilibrium distribution as $f^\chi_{\bm{k}} = n_F\!\boldsymbol{\left(\right.}\!\!\epsilon^\chi(\bm{k})\!\!\boldsymbol{\left.\right)} + g^\chi_{\bm{k}}$,
where $n_F$ is the Fermi-Dirac distribution function and $g^\chi_{\bm{k}}$ is the deviation from equilibrium. Since we are interested in the linear response to weak electric fields $\bm{E}$, we expand $g^\chi_{\bm{k}}$ to linear order in $\bm{E}$. Using Eqs.\ \eqref{def_EOM_r} and \eqref{def_EOM_k}, the Boltzmann equation can then be rewritten to linear order in $\bm{E}$ as
\begin{align}
\label{def_linearBE}
&{-}e\, D^\chi(\bm{k})^{-1}\, \bigg( \frac{\partial n_F}{\partial \epsilon}\bigg\rvert_{\epsilon=\mu}\,\bm{E} \cdot \left[\bm{v}^\chi_{\bm{k}}
  + e\bm{B}\left(\bm{\Omega}^\chi_{\bm{k}} \cdot \bm{v}^\chi_{\bm{k}}\right)\right] \nonumber \\
&\quad{} + \left[\bm{v}^\chi_{\bm{k}} \times \bm{B}\right] \cdot \bm{\nabla}_{\bm{k}}g^\chi_{\bm{k}} \bigg)
  = \sum_{\chi^\prime,\bm{k}^\prime} W_{\bm{k}\bm{k}^\prime}^{\chi\chi^\prime}
  \left(g^{\chi\prime}_{\bm{k}^\prime}-g^\chi_{\bm{k}}\right),
\end{align}
with $D^\chi(\bm{k})=1+e\bm{B}\cdot\bm{\Omega}^\chi_{\bm{k}}$. Energy conservation restricts the sum over $\bm{k}^\prime$ to states with $\epsilon^\chi(\bm{k})=\epsilon^{\chi^\prime}(\bm{k}^\prime) =\mu$. We solve Eq.\ \eqref{def_linearBE} by rewriting $g^\chi_{\bm{k}}$ as~\cite{Sondheimer1962,Taylor1963}
\begin{equation}
\label{ansatz_deviationfunc}
	g^\chi_{\bm{k}} = e\, \frac{\partial n_F}{\partial \epsilon}\bigg\rvert_{\epsilon=\mu}\, \bm{E}\cdot\bm{\Lambda}^\chi_{\bm{k}}, 
\end{equation}
where $\bm{\Lambda}^\chi_{\bm{k}}$ is the vector mean free path, which yields three decoupled equations for the components $\Lambda^\chi_{i,\bm{k}}$, 
\begin{align}
\label{eq_vectormfp}
& D^\chi(\bm{k})^{-1} \left[v^\chi_{i,\bm{k}}
	+ eB_i \left(\bm{\Omega}^\chi_{\bm{k}} \cdot \bm{v}^\chi_{\bm{k}}\right)
	+ e\left(\bm{v}^\chi_{\bm{k}} \times \bm{B}\right) \cdot \bm{\nabla}_{\bm{k}} \Lambda^\chi_{i,\bm{k}}\right]
	\nonumber \\
&\quad {} =- \sum_{\chi^\prime,\bm{k}^\prime} W_{\bm{k}\bm{k}^\prime}^{\chi\chi^\prime}
\left(\Lambda^{\chi\prime}_{i,\bm{k}^\prime}-\Lambda^\chi_{i,\bm{k}}\right).
\end{align}
We choose the $z$ axis to be parallel to $\bm{B}$ and focus on the longitudinal magnetoconductance for $\bm{E}\parallel \bm{B}$, in which case only $\Lambda^\chi_{z,\bm{k}}$ is relevant. Equation \eqref{eq_vectormfp} can be solved with the ansatz
$\Lambda^\chi_{z,\bm{k}}\equiv\Lambda^\chi_{\mu}(\theta)$, which only depends on the polar angle $\theta$ and, via the modulus $k$, on the chemical potential $\mu$, but not on the azimuthal angle $\phi$. This reflects the rotational symmetry of the system (Fermi surfaces, scattering mechanism, and applied fields) around the $k_z$ axis. This ansatz is self-consistent: The group velocity being independent of $\phi$ implies the first two terms on the left-hand side of Eq.\ \eqref{eq_vectormfp} to be independent of $\phi$ as well. The third term vanishes for our ansatz. In the scattering rate, finally, the spinor overlap is
\begin{align}
\label{spinor_overlap}
&\left| \big< u^{\chi^\prime}(\bm{k}^\prime)\big| \hat{V}^{\chi\chi^\prime}_{\bm{k}\bm{k}^\prime}
  \big| u^{\chi}(\bm{k})\big> \right|^2
  = \frac{1}{2}\, \bigl|V^{\chi\chi^\prime} \bigr|^2\, \nonumber\\
  &\times\big( 1+ \chi\chi^\prime
\left[\cos\theta\cos\theta^\prime+\sin\theta\sin\theta^\prime\cos(\phi-\phi^\prime)\right] \big).
\end{align}
The term containing $\phi-\phi^\prime$ vanishes under the sum over $\bm{k}^\prime$ for $\phi$-independent $\Lambda^\chi_{z,\bm{k}}$.

We evaluate the scattering integral in the continuum limit, for which a modification of the density of states induced by the Berry phase has to be taken into account, $\sum_{\bm{k}^\prime} \cdots \rightarrow \mathcal{V} \int d^3k^\prime/(2\pi)^3\: D^\chi(\bm{k}) \cdots$ \cite{Xiao2005}. Equation \eqref{eq_vectormfp} then becomes 
\begin{equation}
\label{eq_lambda_mu}
h^{\chi}_{\mu}(\theta) - \frac{\Lambda^\chi_{\mu}(\theta)}{\tau^\chi_\mu(\theta)}
  = - \sum_{\chi^\prime} \mathcal{V} \int \frac{d^3k^\prime}{(2\pi)^3}\, D^{\chi^\prime}(\bm{k}^\prime)\,
  W_{\bm{k}\bm{k}^\prime}^{\chi\chi^\prime} \Lambda^{\chi\prime}_{\mu}(\theta^\prime),
\end{equation}
with 
\begin{align}
	\frac{1}{\tau^\chi_{\mu}(\theta)}
	&= \sum_{\chi^\prime}\mathcal{V} \int \frac{d^3k^\prime}{(2\pi)^3}\, D^{\chi^\prime}(\bm{k}^\prime)\,
	W_{\bm{k}\bm{k}^\prime}^{\chi\chi^\prime}, \\
	h^{\chi}_{\mu}(\theta)
	&= D^\chi(\bm{k})^{-1} \left[v^\chi_{z,\bm{k}}
	+ eB \left(\bm{\Omega}^\chi_{\bm{k}}\cdot\bm{v}^\chi_{\bm{k}}\right)\right].	
\end{align}
For elastic scattering, the integral over the full momentum space can be replaced by an integral over a surface of constant energy,
\begin{align}
&\sum_{\chi^\prime} \mathcal{V} \int \frac{d^3k^\prime}{(2\pi)^3}\,
	D^{\chi^\prime}(\bm{k}^\prime)\,
	W_{\bm{k}\bm{k}^\prime}^{\chi\chi^\prime} \Lambda^{\chi\prime}_{\mu}(\theta^\prime) \nonumber\\
&~ {}= \sum_{\chi^\prime} \mathcal{V} \int \frac{d\epsilon\, d\theta^\prime\, d\phi^\prime}{(2\pi)^3}\,
	\frac{(k^{\chi^\prime})^3 \sin\theta^\prime}
	{|\bm{v}^{\chi^\prime}_{\bm{k}^\prime} \cdot \bm{k}^\prime|}\,
	D^{\chi^\prime}(\bm{k}^\prime)\, W_{\bm{k}\bm{k}^\prime}^{\chi\chi^\prime}
	\Lambda^{\chi\prime}_{\mu}(\theta^\prime) \nonumber\\
&~ {}= \sum_{\chi^\prime} \frac{n}{4\pi} \int d\theta^\prime \sin\theta^\prime\,
	\frac{(k^{\chi^\prime})^3}{|\bm{v}^{\chi^\prime}_{\bm{k}^\prime} \cdot \bm{k}^\prime|}\,
	D^{\chi^\prime}(\bm{k}^\prime)\, \bigl|V^{\chi\chi^\prime} \bigr|^2 \nonumber \\
&~\quad{} \times \left( 1 + \chi\chi^\prime \cos\theta\cos\theta^\prime \right)
	\Lambda^{\chi\prime}_{\mu}(\theta^\prime) .
\end{align}
We may hence rewrite Eq.\ \eqref{eq_lambda_mu} as
\begin{align}
\label{eq_lambdamu}
&h^{\chi}_{\mu}(\theta) - \frac{\Lambda^\chi_{\mu}(\theta)}{\tau^\chi_\mu(\theta)}
	=- \sum_{\chi^\prime} \frac{n}{4\pi} \int d\theta^\prime \sin\theta^\prime\,
	\frac{(k^{\chi^\prime})^3}{|\bm{v}^{\chi^\prime}_{\bm{k}^\prime} \cdot \bm{k}^\prime|} \nonumber \\
&\quad{} \times D^{\chi^\prime}(\bm{k}^\prime)\, \bigl|V^{\chi\chi^\prime} \bigr|^2
	\left( 1 + \chi\chi^\prime \cos\theta\cos\theta^\prime \right)
	\Lambda^{\chi\prime}_{\mu}(\theta^\prime) . 
\end{align}
We now make the ansatz $\Lambda^\chi_{\mu}(\theta) = -\tau^{\chi}_\mu(\theta)
[-h^{\chi}_{\mu}(\theta)+\lambda^\chi + \chi\delta^\chi\cos\theta]$,
which leads to a system of coupled linear equations for the four real coefficients $\lambda^+$, $\lambda^-$, $\delta^+$, and $\delta^-$. Details on their solutions are given in the Supplemental Material \cite{Note1}. Inserting the solution for $\Lambda^\chi_{z,\bm{k}}$ into Eq.\ \eqref{ansatz_deviationfunc}, we obtain the current density
\begin{equation}
\label{def_currentdens}
	\bm{j}=-\frac{e}{\mathcal{V}}\sum_{\chi,\bm{k}}\dot{\bm{r}}^\chi f^\chi_{\bm{k}}. 
\end{equation}

\begin{figure}
		\centering
		\includegraphics[width=0.95\columnwidth]{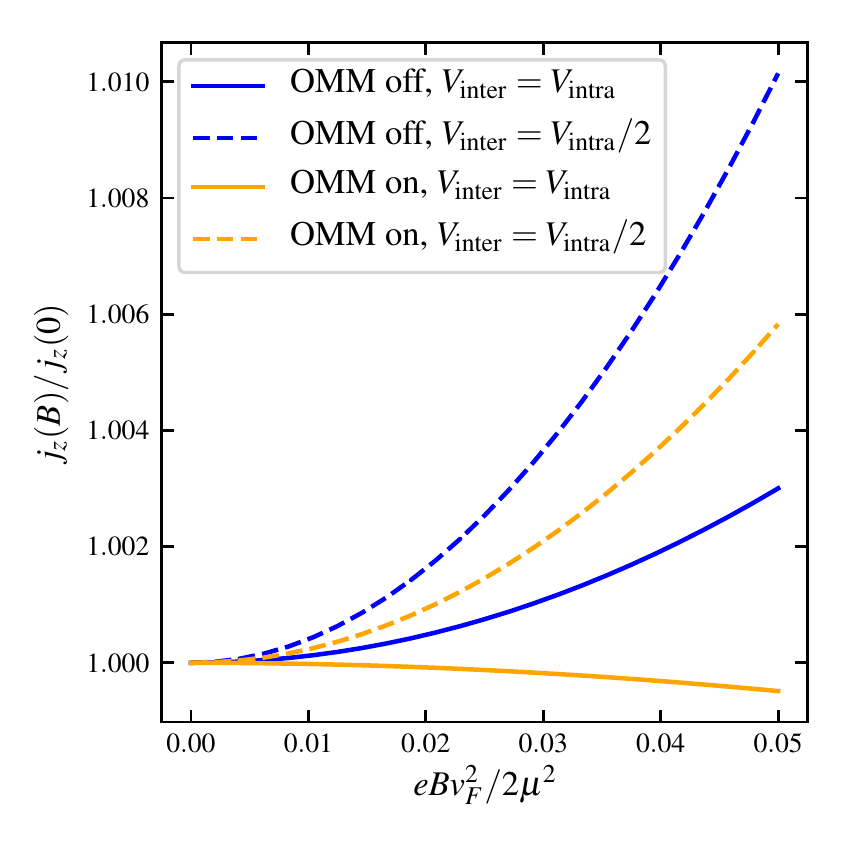}
		\caption{Current density $j_z(B)/j_z(0)$ as a function of $eBv_F^2/2\mu^2$ in the presence (orange curves) and absence (blue curves) of the OMM. For the solid curves, the intervalley scattering amplitude is equal to the intravalley scattering amplitude, $V_\mathrm{intra}=V_\mathrm{inter}=3$, whereas for the dashed curves, $V_\mathrm{inter}=1.5$ is half as large as $V_\mathrm{intra}=3$.}  
		\label{Fig_2}
\end{figure}

\textit{Results}. We study the magnetoconductance associated with an electric field $\bm{E}=E\bm{e}_z\parallel \bm{B}$ at zero temperature. As discussed above, the rotational symmetry of the system around the $z$ axis forces the resulting current to flow along the $z$ direction, $\bm{j}=j_z\bm{e}_z$. In Fig.\ \ref{Fig_2}, the current density $j_z(B)/j_z(0)$ is plotted with and without taking the OMM into account, and for $V_\mathrm{inter}=V_\mathrm{intra}$ and $V_\mathrm{inter}=V_\mathrm{intra}/2$.

In agreement with earlier work, we obtain a positive magnetoconductance if the OMM is neglected \cite{SpivakAndreev2016,SonSpivak2013,DasAgarwal2019,Imran2018,Kim2014,Dantas2018,Johansson2019,Burkov2014,PesinMishchenko2015}. Decreasing the intervalley scattering amplitude relative to the intravalley scattering amplitude leads to an increase of the magnetoconductance. This is plausible since weak intervalley scattering suppresses the relaxation towards equal occupation of the two valleys and thereby enhances signatures of the chiral anomaly. In agreement with this result, Son and Spivak \cite{SonSpivak2013}, as well as Kim \textit{et al.}\ \cite{Kim2014}, have found positive magnetoconductance for the limit of dominant intravalley scattering, $V_\mathrm{inter}\ll V_\mathrm{intra}$. We here find that the positive magnetoconductance proportional to $\bm{B}^2$ persists all the way to dominant short-range scattering in the absence of OMM. This is in accord with Burkov \cite{Burkov2014}, who considered Gaussian impurities in quantum field terminology, and with Johansson \textit{et al.}\ \cite{Johansson2019}. 

Including the OMM dramatically changes this physics, as shown by the two orange curves in Fig.\ \ref{Fig_2}. For equal intra- and intervalley scattering amplitudes, we find a \emph{negative} magnetoconductance. Only when the intervalley scattering amplitude is reduced, the redistribution of charges between the two valleys is enhanced, such that the magnetoconductance eventually changes sign. Figure \ref{Fig_2} illustrated this behavior for $V_\mathrm{inter}=V_\mathrm{intra}/2$, in which case we find the positive magnetoconductance known from the limit $V_\mathrm{inter}\ll V_\mathrm{intra}$ \cite{DasAgarwal2019OMM,GrushinVenderbos2016,Cortijo2016}. For the present model, the sign change occurs for
\begin{equation}
\label{ratio_scattering_amplitudes}
\frac{V_\mathrm{inter}}{V_\mathrm{intra}}\approx 0.8825,	
\end{equation}
which is accessible in systems with dominant short-range scattering. Our calculations thus show that the positive sign of the longitudinal magnetoconductance is not universal. For negative longitudinal magnetoconductance in the semiclassical regime, the magnetoconductance in the quantum limit of large $B$ may still have a positive sign depending on the dominant scattering mechanism \cite{LiRoyDasSarma2016,Goswami2015,Zhang2016,Goswami2015,Shao2019,JiLu2018}. This would imply the existence of a critical magnetic field $B^*$ at which the magnetoconductance changes sign.

To trace the origin of the sign change of the magnetoconductance upon including the OMM, we attach a factor $\eta_B\in\{0,1\}$ to the Berry curvature in Eq.\ \eqref{def_Berrycurve}, and a factor $\eta_O\in\{0,1\}$ to the OMM in Eq.\ \eqref{def_OMM}. These factors allow us to independently switch on and off the effects of the Berry curvature and the OMM. Next, we expand the expression for the current density $j_z$ to second order in the magnetic field. For equal intra- and intervalley scattering amplitudes, and using $\eta_i^2=\eta_i$, we find 
\begin{equation}
\label{eq_jz_Bsquare}
j_z(B) = \frac{v_Fe^2 E}{\pi n V_\mathrm{intra}^2}\, \bigg(
  \frac{1}{3} + \frac{\alpha^2}{45}\, [ 18\, \eta_B + 11 \,\eta_O - 32\, \eta_B\, \eta_O ] \bigg),
\end{equation} 
with $\alpha=eBv_F^2/2\mu^2$. Evidently, the sign change of the magnetoconductance is not the result of a simple competition between the Berry curvature and the OMM---both mechanisms separately favor a positive magnetoconductance. Only their concurrence leads to a sign change since the third term in the angular brackets in Eq.\ \eqref{eq_jz_Bsquare} outweighs the first two. Our findings motivate future studies of the interplay of these two effects, also in relation to additional scattering mechanisms such as Coulomb or magnetic scattering.

\textit{Summary and conclusions}. We have investigated the longitudinal magnetoconductance for the generic low-energy model of a Weyl semimetal consisting of two isotropic Weyl nodes of opposite chirality in homogeneous parallel electric and magnetic fields, including both the Berry curvature and the OMM of wave packets in the semiclassical limit. If the OMM is neglected, we find a positive longitudinal magnetoconductance quadratic in the magnetic field even in the limit of pure $\delta$-function scattering. In stark contrast, the inclusion of the OMM leads to a \emph{negative} magnetoconductance quadratic in field if intervalley scattering is sufficiently strong. Previous transport measurements on Weyl semimetals that exhibited a negative magnetoconductance in some regimes \cite{Huang2015_b,Li2016,Zhang2016_b,Naumann2020} should be revisited and new experiments performed with our mechanism in mind.

We have traced the sign reversal of the longitudinal magnetoconductance to the concurrence of the OMM and the Berry curvature. Hence, \emph{both} effects must be included to correctly describe transport in Weyl semimetals in a weak magnetic field.
A sign reversal of the \emph{transverse} (perpendicular to $\bm{B}$) magnetoconductance induced by the OMM has been discussed by Das and Agarwal \cite{DasAgarwal2019OMM}, for tilted nodes. Unlike our study, Ref.\ \cite{DasAgarwal2019OMM} does not report a sign change of the longitudinal magnetoconductance, possibly due to relaxation-time approximation used there, compared to our exact solution of the Boltzmann equation.
Reference \cite{Xiao2020} shows, in its recent versions, that the sign change in the longitudinal magnetoconductance due to the OMM and intrascattering effects is not limited to Weyl systems. In Ref.~\cite{Xiao2020}, a two-dimensional gapped Dirac model is considered including the OMM, intrascattering effects, and the side-jump effect, which have not been considered in this work.

Our exact solution of the Boltzmann equation shows that the effect of the OMM on transport is in general much more important, and can lead to a quadratic negative magnetoconductance akin to the one of simple metals \cite{Kohler1938}. Our results suggest that the deduction of Weyl physics from magnetotransport requires a detailed modeling of the microscopic scattering mechanisms in a given material, and underlines that magnetotransport is most sensitive to Weyl physics in systems with dominant long-range scattering.

\textit{Acknowledgments}. We are grateful for helpful discussions with A.\ Johansson, M.\ Breitkreiz, and A.\ Zyuzin. Financial support by the Deut\-sche For\-schungs\-ge\-mein\-schaft via the Emmy Noether Programme ME4844/1-1, the Collaborative Research Center SFB 1143, project A04, and the Cluster of Excellence on Complexity and Topology in Quantum Matter ct.qmat (EXC 2147) is acknowledged.

\bibliography{Knoll}
\onecolumngrid

\newpage

\renewcommand{\theequation}{S\arabic{equation}}
\renewcommand{\thefigure}{S\arabic{figure}}
\setcounter{page}{1}
\setcounter{equation}{0}
\setcounter{figure}{0}

\setcounter{secnumdepth}{3}

\begin{center}
	\textbf{{\large Supplemental Material for\\[0.5ex]
			Negative longitudinal magnetoconductance at weak fields in Weyl semimetals}}\\[1.5ex]
	Andy Knoll, Carsten Timm, and Tobias Meng
\end{center}

\section{Semiclassical limit}

In this section, we derive the condition for the magnetic field range in which in the semiclassical limit is justified. For a magnetic field along the $k_z$-direction, the dispersion of the $m$-th Landau level of positive energy with $m>0$ is given by 
\begin{equation}
\epsilon_m (k_z) = v_F\sqrt{2eBm+(v_F k_z)^2}.	
\end{equation}
The number of occupied Landau levels must be large in the semiclassical limit, $n\gg 1$. Equivalently, the energy splitting
\begin{equation}
\label{eq_energysplitting}
\Delta\epsilon(B) \equiv v_F\sqrt{2eB(n+1)}-v_F\sqrt{2eB n}
\end{equation}
between the last occupied and the first unoccupied Landau level for $k_z = 0$ should be small compared to the chemical potential,
\begin{equation}
\label{condition_smallBfield}
\Delta\epsilon(B)\ll \mu.
\end{equation} 
The energy splitting can be estimated as
\begin{equation}
\Delta\epsilon(B)
= v_F \sqrt{2eBn} \left(\sqrt{1+\frac{1}{n}}-1\right)
\cong v_F\sqrt{2eBn}\left(1+\frac{1}{2n}-1\right)
= v_F \sqrt{\frac{eB}{2n}}
\label{eq_estimation_energysplit}.
\end{equation}
Since $n$ is the index of the last occupied Landau level, we have
\begin{equation}
\label{ineq_chempot_LL}
v_F\sqrt{2eBn}<\mu<v_F\sqrt{2eB(n+1)}.
\end{equation}
Using Eq.~\eqref{ineq_chempot_LL}, we obtain
\begin{equation}
\label{eq_n_terms_of_B}
n= \left\lfloor \frac{1}{2eB}\left(\frac{\mu}{v_F}\right)^2 \right\rfloor ,
\end{equation}
where $\lfloor x\rfloor$ is the largest integer smaller or equal to $x$. By combining Eqs.\ \eqref{condition_smallBfield}, \eqref{eq_estimation_energysplit}, and \eqref{eq_n_terms_of_B}, we obtain the condition
\begin{equation}
B \ll \frac{1}{e}\left(\frac{\mu}{v_F}\right)^{\!2}
\end{equation}
for the magnetic field to be considered weak and the semiclassical approximation to be valid.

\section{Determination of the coefficients \texorpdfstring{$\lambda^\chi$}{} and \texorpdfstring{$\delta^\chi$}{}}
\begin{figure}
	\centering
	\includegraphics{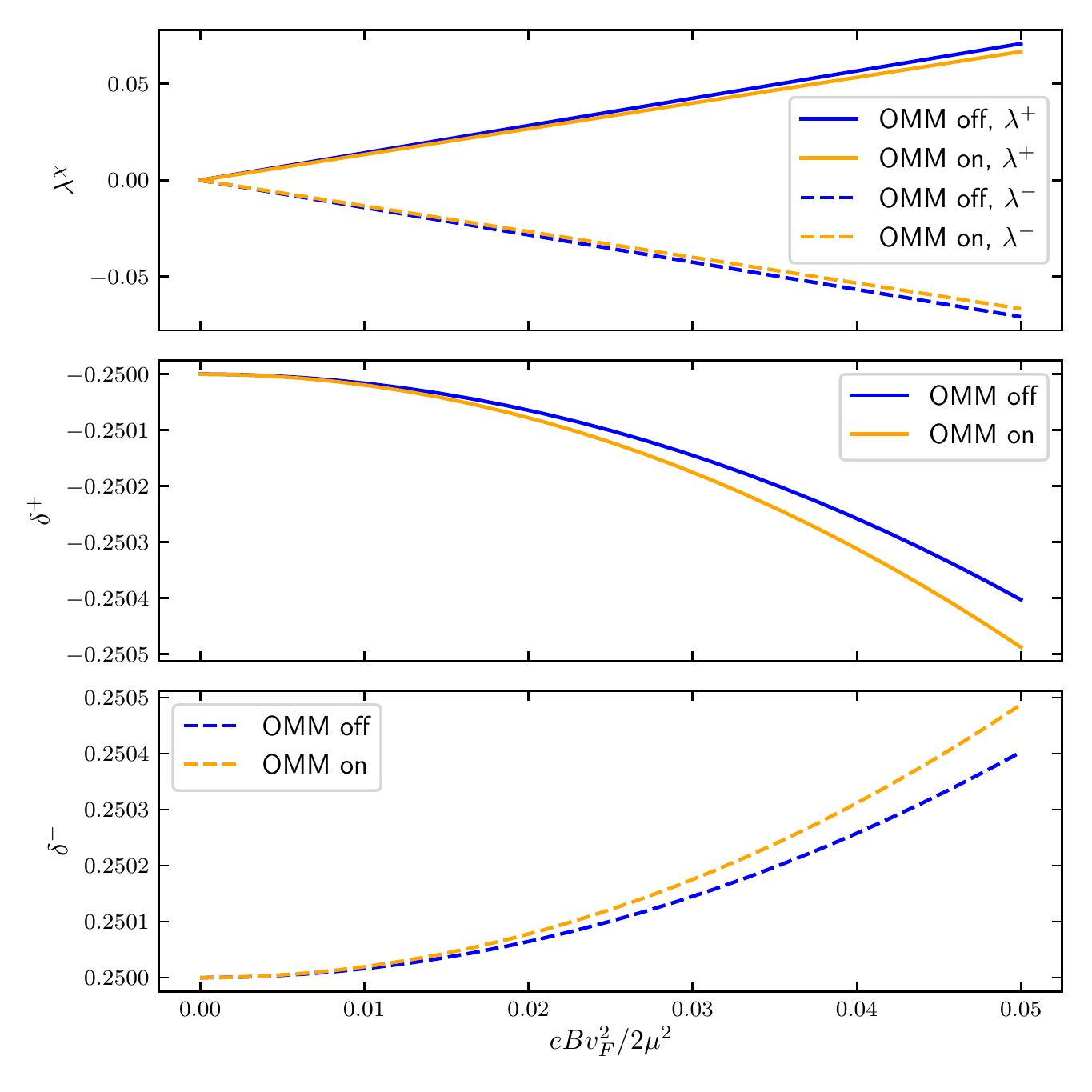}
	\caption{Coefficients $\lambda^\chi$ and $\delta^\chi$ in the presence and the absence of the OMM for $V_\mathrm{inter} = V_\mathrm{intra}/2$.}
	\label{figS1}
\end{figure}

The ansatz for the vector mean free path given in the main text,
\begin{equation}
\label{S.Ansatz_lambdamu}
\Lambda^\chi_{\mu}(\theta)
= - \tau^{\chi}_\mu(\theta)\left(-h^{\chi}_{\mu}(\theta)+\lambda^\chi+ \chi\delta^\chi\cos\theta\right),
\end{equation}
contains the four real coefficients $\lambda^\chi$ and $\delta^\chi$. Recall that $\chi=\pm$ denotes the chirality of the Weyl node. In this section, we present details on their determination, as there is a subtlety. We have to solve the equation
\begin{equation}
\label{eq_vectormeanfreepath}
h^{\chi}_{\mu}(\theta) - \frac{\Lambda^\chi_{\mu}(\theta)}{\tau^\chi_\mu(\theta)}
= - \sum_{\chi^\prime} \frac{n}{4\pi} \int d\theta^\prime \sin\theta^\prime\,
\frac{(k^{\chi^\prime})^3}{|\bm{v}^{\chi^\prime}_{\bm{k}^\prime} \cdot \bm{k}^\prime|}\,
D^{\chi^\prime}(\bm{k}^\prime)\, \bigl|V^{\chi\chi^\prime} \bigr|^2
\left( 1 + \chi\chi^\prime \cos\theta\cos\theta^\prime \right)
\Lambda^{\chi\prime}_{\mu}(\theta^\prime) . 
\end{equation}
By inserting Eq.\ \eqref{S.Ansatz_lambdamu}, we obtain a system of equations for the four coefficients $\lambda^+$, $\lambda^-$, $\delta^+$, and $\delta^-$:
\begin{equation}
\label{eq_matrix_coeffs}
\begin{pmatrix}R_1^+\\R_1^-\\R_2^+\\R_2^-\end{pmatrix}=
\begin{pmatrix}C_1^{++}-1 & C_1^{+-} & C_2^{++} & C_2^{+-} \\
C_1^{-+} & C_1^{--}-1 & C_2^{-+} & C_2^{--} \\
C_2^{++} & C_2^{+-} & C_3^{++}-1 & C_3^{+-} \\
C_2^{-+} & C_2^{--} & C_3^{-+} & C_3^{--}-1 \end{pmatrix}\begin{pmatrix}\lambda^+\\\lambda^-\\\delta^+\\\delta^-\end{pmatrix},
\end{equation}
where 
\begin{align}
p^{\chi\chi^\prime}(\theta) &= \frac{n}{4\pi}\, \sin\theta\,
\frac{(k^{\chi^\prime})^3}{|\bm{v}^{\chi^\prime}_{\bm{k}} \cdot \bm{k}^\prime|}\,
D^{\chi^\prime}(\bm{k})\, \big|V^{\chi\chi^\prime}\big|^2, \\
R_1^{\chi} &= \sum_{\chi^\prime} \int d\theta^\prime\, p^{\chi\chi^\prime}(\theta^\prime)\,
h_\mu^{\chi^\prime}(\theta^\prime),\\
R_2^{\chi} &= \sum_{\chi^\prime} \int d\theta^\prime\, p^{\chi\chi^\prime}(\theta^\prime)\,
\chi^\prime\cos\theta^\prime\, h_\mu^{\chi^\prime}(\theta^\prime),\\
C_1^{\chi\chi^\prime} &= \int d\theta^\prime\, p^{\chi\chi^\prime}(\theta^\prime),\\
C_2^{\chi\chi^\prime} &= \int d\theta^\prime\, p^{\chi\chi^\prime}(\theta^\prime)\,
\chi^\prime\cos\theta^\prime,\\
C_3^{\chi\chi^\prime} &= \int d\theta^\prime\, p^{\chi\chi^\prime}(\theta^\prime)\,
\cos^2\theta^\prime.
\end{align}
Explicit evaluation shows that the coefficient matrix in Eq.~\eqref{eq_matrix_coeffs} has rank $3$. Consequently, it has a one-parameter family of solutions. The origin of this apparent arbitrariness is that the solution of the \emph{linearized} Boltzmann equation and hence of Eq.~\eqref{eq_vectormeanfreepath} is only determined up to a constant: if $\Lambda^\chi_{\mu}(\theta)$ solves Eq.~\eqref{eq_vectormeanfreepath} then $\Lambda^\chi_{\mu}(\theta)+c$ with $c$ an arbitrary constant does so as well. The physical solution is found by imposing electron-number conservation,
\begin{equation}
\label{eq_norm_cond}
\sum_{\chi,\bm{k}} g^\chi_{\bm{k}}=0.
\end{equation}
By solving Eqs.~\eqref{eq_vectormeanfreepath} and \eqref{eq_norm_cond} simultaneously we obtain the results given in the main text. For completeness, the coefficients $\lambda^\chi$ and $\delta^\chi$ are plotted in Fig.~\ref{figS1} as functions of $\alpha\,=\,eBv_F^2/2\mu^2$ for the dotted curves in Fig.~2 ($V_\mathrm{inter}=V_\mathrm{intra}/2$) of the main text. For the solid curves in Fig.~2 ($V_\mathrm{inter}=V_\mathrm{intra}$) the four coefficients $\lambda^\chi$ and $\delta^\chi$ vanish.

\end{document}